# Web Document Analysis for Companies Listed in Bursa Malaysia

Mohd Shahizan Othman and Lizawati Mi Yusuf

Faculty of Computer Science and Information System,
Universiti Teknologi Malaysia, 81310 Skudai,
Johor, Malaysia.
{shahizan, lizawati}@utm.my

Juhana Salim

Faculty of Information Science and Technology,
Universiti Kebangsaan Malaysia,
43600 Bangi, Selangor,
js@ftsm.ukm.my

*Abstract*- **This paper discusses a research on web document analysis for companies listed on Bursa Malaysia which is the forerunner of financial and investment center in Malaysia. Data set used in this research are from the company web documents listed in the Main Board and Second Board on Bursa Malaysia. This research has used the Web Resources Extraction System which was developed by the research group mainly to extract information for the web documents involved. Our research findings have shown that the level of website usage among the companies on Bursa Malaysia is still minimal. Furthermore, research has also found that 60.02% of the image files are utilised making it the most used type of file in creating websites.**

## I. INTRODUCTION

Accessible Internet services have significanty increased the number of websites. Among the researches done, it was discovered that approximately one billion websites exist and 1.5 million brand-new websites are produced every day. This amount has escalated and has reached 11.5 billion websites as of January 2005 [1, 2, 3]. Internet domains have also expanded drastically. According to the Internet Systems Consortium [4], in January 1993, the total of domains were 1,313,000. However, in January 2008, the amount has already risen up to 541,677,360. Apart from that, the global growth of the Internet has impacted Internet usage in Malaysia. A research conducted by the Internet World Star [5] has shown that the growth of the Internet usage in Malaysia has significantly increased. In 1995, there were only 25,000 Internet users in Malaysia and the amount has multiplied to 14.9 million in 2007. Hence, this research paper will discuss on the prospects of website usage amongst the listed companies on Bursa Malaysia. The discussion is divided into four sections which are web documents, data set, methodology and research findings.

## II. WEB DOCUMENT

Today, web documents exist in different forms and structures. Web documents can be written using different types of scripting languages such as HyperText Markup Language (HTML), Active Server Pages (ASP) and HyperText Preprocessor (PHP). Although there are various types of scripting languages that can be used to create web documents, the final output of the source code is rather similar. Based on a research by Lyman and Hal [6], it is shown that a majority of 30.8% of the web documents consist of HTM/HTML and PHP file. Other files used to create web documents are image files with 23.2%, Adobe with 9.2%, animation with 4.3%, compressed file with 3.8% and others with 28.7% (refer Figure 1).

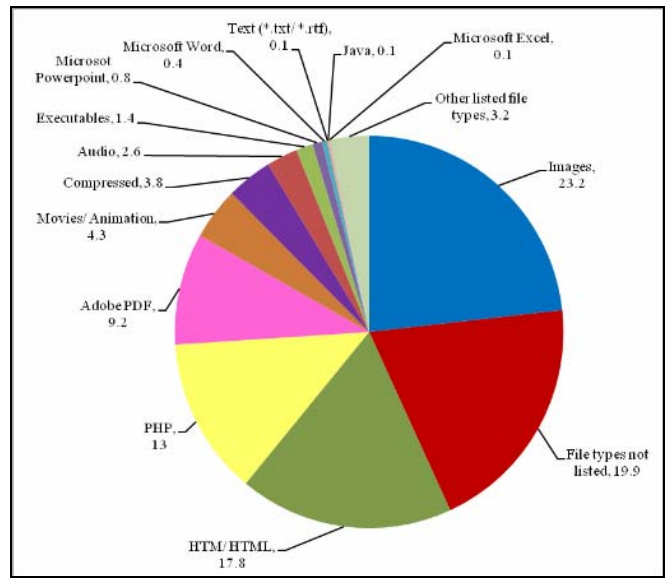

Figure 1. Percentage of web documents according to types of files [6]

According to the founder of the World Wide Web (WWW), Berners-Lee [7], webs are catered for everyone. If the users wish to create a web document, it is crucial for everyone to be able to understand and experience a connection to it. Generally, HTML documents consist of meta tag, tag and content [8]. Web documents are made up of syntax documents which explain the structure, types of presentations, semantics or any external reactions (refer Figure 2) [9]. Web document syntax consists of a mixture of tags which creates a complete web process command. The research on web document structure by Etzioni [10], Kosala and Blockheel [11] had found that most of the webs emphasized on the presentation style of web document on the server compared to its content.

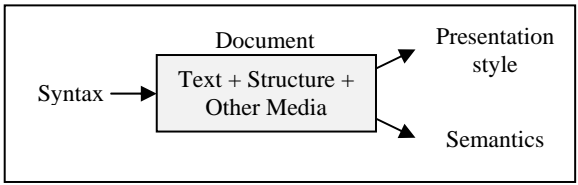

Figure 2. Characteristics of a document [8]

Tags are text used in each HTML document to explain types of elements, format, structure and document layout. There are 109 HTML tags which exist. However, all of the texts will not be used simultaneously in creating a web document. The <html> tag is the primary tag in a web document. This tag will inform web servers that the document is an HTML document. Without this tag, web servers will assume the web document as a usual text document. Tags are divided into two which are <head> and <body> tag. The <head> tag attributes bibliography in certain HTML documents. On the other hand, <body> tag is used to insert contents in web documents. All texts in <body> tag will be manipulated to HTML and will be screened on the web server.

### III. DATASET

Data set used in this research are company data listed on Bursa Malaysia. Bursa Malaysia is a forerunner of financial and investment center in Malaysia. It is divided into three which are the Main Board, Second Board and the Mesdaq Market. The Main Board consists of 658 listed companies, followed by 237 companies in Second Board and 69 companies in Mesdaq Market, respectively. Apart from that, this research only utilises company data in the Main Board and Second Board as the research data set. Listed companies on Bursa Malaysia can also be categorised based on types of businesses conducted such as consumer products, industrial products and contructions. Table I shows a data set on Bursa Malaysia.

TABLE I

Number of companies on Bursa Malaysia with websites according to respective categories

| Code | Category | Main Board | Second Board | Total |
|------|----------|------------|--------------|-------|
| IP | Industrial Products | 154 | 130 | 284 |
| TS | Trading and Services | 138 | 48 | 186 |
| CP | Consumer Products | 93 | 50 | 143 |
| P | Properties | 98 | 3 | 101 |
| C | Constructions | 43 | 16 | 59 |
| F | Finance | 48 | 0 | 48 |
| Pl | Plantations | 42 | 4 | 46 |
| T | Technology | 17 | 6 | 23 |
| I | Infrastructure | 9 | 0 | 9 |
| R | Reits | 7 | 0 | 7 |
| H | Hotels | 5 | 0 | 5 |
| CE | Close-End Fund | 2 | 0 | 2 |
| E | Exchange Traded Funds | 1 | 0 | 1 |
| M | Mining | 1 | 0 | 1 |
| | **Total** | **658** | **257** | **915** |

### IV. RESEARCH METHODOLOGY

In this research we have developed Web Resource Extraction System (WRES) software to simplify the process of extracting web documents. Extracting process starts with the WRES program that reads the web document. Secondly, the document processing will take place where HTML script and tag is removed. Next, while the document processing is going on, WRES program will communicate with the database module to retrieve stopwords and to restore independent files from any tags, scripts and stopwords. This process will continue until all documents undergo document processing (refer Figure 3).

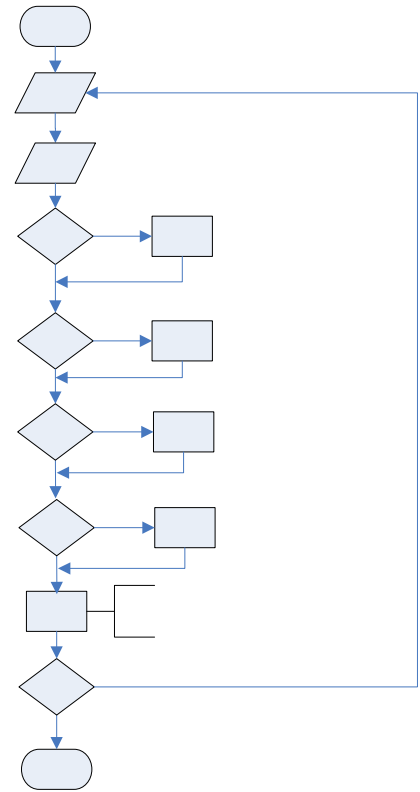

Figure 3. Flowchart Web Resource Extraction System

### V. RESEARCH FINDINGS

Based on the research conducted, it is found that website usage amongst the listed companies on Bursa Malaysia is still minimal. Only 64.53% of the companies have websites. However, not all operating websites can be used. This happens because there are some websites that have problems such as wrong server, non-existing websites and also image websites. There are 41 companies and sister companies that share the same websites. For example, Berjaya Corporation Bhd, Cosway Coporation Bhd, Berjaya Capital Bhd and Matrix International Bhd that share http://www.berjaya.com.

Therefore, this research has discovered that only 52.95% from the total of companies have websites listed in Bursa

Malaysia Board. Figure 4 shows the 72.34% of companies listed in the Main Board have websites, followed by 20.92% from the Second Board with 6.7% from Mesdaq Market.

In addition, this research had studied company websites at the Main Board and Second Board according to category. The highest percentage of companies that have websites according to category is in the Industrial Products (31.04%), followed by 20.33% in the Products and Services, 15.63% in the Consumer Goods and 11.04% in Land Properties. Less than 10% of companies listed in 10 categories have websites, they include: Construction, Finance, Plantation, Techonology, Reits, Hotel, Infrastructure, Closed-End Fund, Exchangeable Goods and Mining Fund.

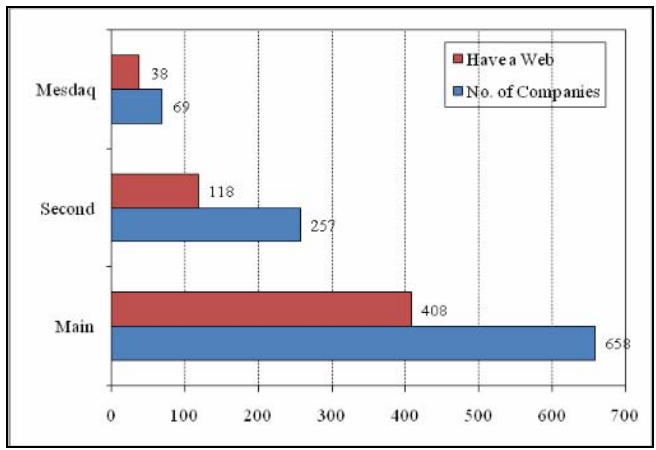

Figure 4. Companies with websites listed on Bursa Malaysia

The research on companies with websites on the Main Board and the Second Board according to fields showed that the Industrial Goods recorded the highest percentage of 17.05% followed by Goods and Services with 13.33% and Consumers Goods with 8.19%. There are no companies in the Reits and Mining field that have websites. Meanwhile, for each Closed-Ended Fund and Exchangeable Goods Fund, there is only one website, respectively. Table II depicts detailed information of the company percentage on the Main Board and Second Board on Bursa Malaysia according to fields.

This research findings have shown to be in the favour of the research findings collected by Ruhaya *et al.* [12], Noor Azizi and Mahamad [13], Rosli *et al.* [14] and also Kamarulbaraini dan Khairul Azman [15] where the percentage of companies on Bursa Malaysia that have websites are still at minimal which is only between 31.6% until 62.0%. Table III shows the percentage of companies on Bursa Malaysia that have websites based on previous researchers.

This research also supports the statistics released by The Star [16] which stated that 87% of the companies in Malaysia do not have websites. According to The Star, the reason behind this is because the companies do not know how to attract Internet users to visit their websites. Below are the discussions in regard to the usage of host domains, usage of type and size file to produce web documents and analyses on the usage of tagging for websites listed on the Main Board and the Second Board of Bursa Malaysia.

TABLE II

Percentage of companies on Main Board and Second Board on Bursa Malaysia

| Code | Main Board | | | Second Board | | | Total |
|------|------|------|------|------|------|------|------|
| | A | B | C | A | B | C | |
| IP | 4.70 | 1.53 | 10.60 | 6.78 | 0.98 | 6.45 | 31.04 |
| TS | 3.39 | 1.20 | 10.49 | 1.97 | 0.44 | 2.84 | 20.33 |
| CP | 3.93 | 0.55 | 5.68 | 2.08 | 0.87 | 2.51 | 15.63 |
| P | 3.50 | 0.98 | 6.23 | 0.33 | 0.00 | 0.00 | 11.04 |
| C | 1.09 | 0.33 | 3.28 | 0.87 | 0.33 | 0.55 | 6.45 |
| F | 1.31 | 0.11 | 3.83 | 0.00 | 0.00 | 0.00 | 5.25 |
| PL | 2.62 | 0.22 | 1.75 | 0.33 | 0.00 | 0.11 | 5.03 |
| T | 0.33 | 0.22 | 1.31 | 0.22 | 0.00 | 0.44 | 2.51 |
| I | 0.11 | 0.00 | 0.87 | 0.00 | 0.00 | 0.00 | 0.98 |
| R | 0.77 | 0.00 | 0.00 | 0.00 | 0.00 | 0.00 | 0.77 |
| H | 0.22 | 0.00 | 0.33 | 0.00 | 0.00 | 0.00 | 0.55 |
| CE | 0.11 | 0.00 | 0.11 | 0.00 | 0.00 | 0.00 | 0.22 |
| E | 0.00 | 0.00 | 0.11 | 0.00 | 0.00 | 0.00 | 0.11 |
| M | 0.11 | 0.00 | 0.00 | 0.00 | 0.00 | 0.00 | 0.11 |
| | 22.19 | 5.14 | 44.59 | 12.57 | 2.62 | 12.90 | 100 |

Reference: A – Companies without websites, B – Companies with difficulties, C – Companies with websites

TABLE III

Research on companies on Bursa Malaysia that have web sites

| Reasearcher(s) | Number of Companies Involved | Year | Companies have Web Sites |
|------|------|------|------|
| Ruhaya *et al.* [12] | 50 | 1999 | 62.0% |
| Noor Azizi and Mahamad [13] | 237 | 1999 | 31.6% |
| Rosli *et al.* [14] | 122 | 2002 | 52.0% |
| Kamarulbaraini and Khairul Azman [15] | 923 | 2004 | 50.9% |

a) Host domains usage

Host domain is essential because it helps users to arrive on websites easily. This research shows that 56.95% of the companies on Bursa Malaysia use ".com.my" host domains, while 38.42% uses ".com" host domains and 4.63% uses others. This means that companies on Bursa Malaysia have the tendency to use the country code domain which is the ".com.my" that shows the company's country compared to using the ".com" global host domain.

b) Usage of type and size file to produce web documents

There are various types of files used to produce layout for the main website of certain companies (refer to Figure 5). Image files such as *.jpg, *.gif, *.png and *.bmp are the most utilised files (60.02%). The usage of image files is needed to produce interesting and user-friendly web document. Furthermore, related files with document such as *.asp, *.cfm, *.doc, *.htm, *.pdf, *.php, *.ppt, *.ps, *.rtf, and *.txt have recorded the highest percentage of 33.36%. Meanwhile, other files recorded 0.96%.

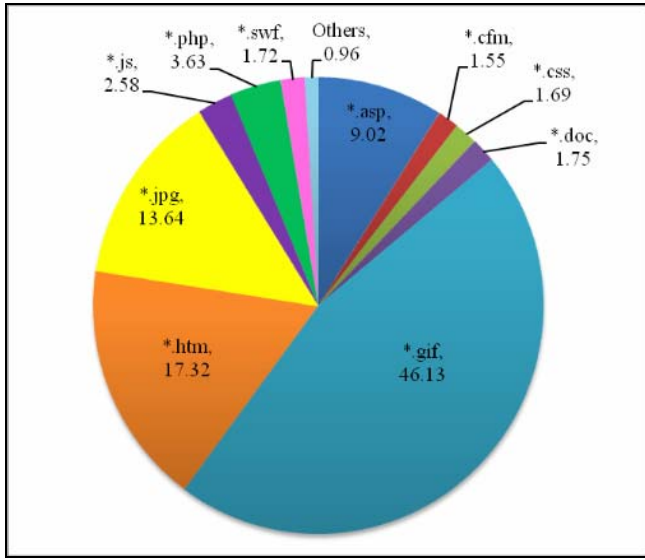

Figure 5. Existing types of files in web documents on Bursa Malaysia

On the other hand, analysis on the file size has shown that image files are the largest in file size where by it represents 62.59% of the total overall size of the web document file where by 38.24MB in *.gif and 37.24 MB in *.jpg file. The size of *.htm and *.txt files is 11.69% and as for the others is 25.72%. Hence, image files represent the highest percentage for the total and file size. Figure 6 depicts the fraction size according to the types of files in companies websites listed in Bursa Malaysia.

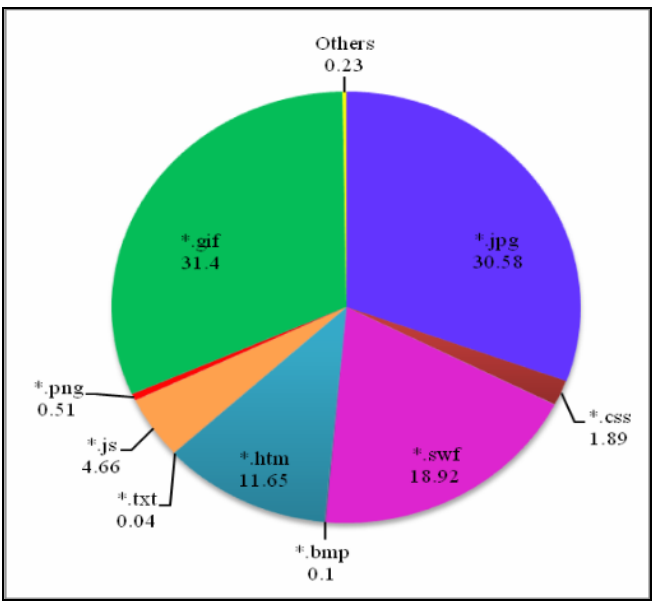

Figure 6. Fraction size according to the types of files on Bursa Malaysia

c) Analysis on Use of Tag

Research on the use of tag was conducted on 37 normally encountered tag in the web document. This research discovered that the average number of tag used is 162.47 per document. The  tag is the most commonly used tag (15.21%) followed by <tr> tag (14.85%) and <a> tag

(10.85%). Our findings showed that in some websites, there are four tags which are <body>, <head>, <title>, and <html> that exist more than once. Those tags usually exist once in a web document. Detailed information pertaining to the use of tag is shown in Table 4.

TABLE IV

Tag Usage

| No | Tag | Total | Average | Percent |
|---|---|---|---|---|
| 1. |  | 12877 | 24.72 | 15.21 |
| 2. | <tr> | 12571 | 24.13 | 14.85 |
| 3. | <a> | 9182 | 17.62 | 10.85 |
| 4. | <font> | 8015 | 15.38 | 9.47 |
| 5. | <td> | 7407 | 14.22 | 8.75 |
| 6. | <br> | 7065 | 13.56 | 8.35 |
| 7. | <div> | 5704 | 10.95 | 6.74 |
| 8. | <table> | 5527 | 10.61 | 6.53 |
| 9. | <!--...--> | 3307 | 6.35 | 3.91 |
| 10. | <p> | 2617 | 5.02 | 3.09 |
| 11. | <b> | 2129 | 4.09 | 2.52 |
| 12. | <strong> | 1378 | 2.64 | 1.63 |
| 13. | <script> | 1306 | 2.51 | 1.54 |
| 14. | <meta> | 1133 | 2.17 | 1.34 |
| 15. | <li> | 922 | 1.77 | 1.09 |
| 16. | <body> | **575** | **1.10** | 0.68 |
| 17. | <head> | **556** | **1.07** | 0.66 |
| 18. | <title> | **547** | **1.05** | 0.65 |
| 19. | <html> | 551 | **1.06** | 0.65 |
| 20. | <style> | 266 | 0.51 | 0.31 |

*Bolded words: Amount of tagging that exceeds its usual amount

This research discovered that there is a significant difference in the percentage of tagging usage for web documents and the average amount of tags per document differs. This difference happen because of the tag that exists only once for each document such as the <head>, <title>, and <body> tag. As for other tags such as <td>, <a>, and <font> tag, they can appear more than once. Figure 7 shows the percentage of tags used in web documents while Figure 8 shows the number of tags per document.

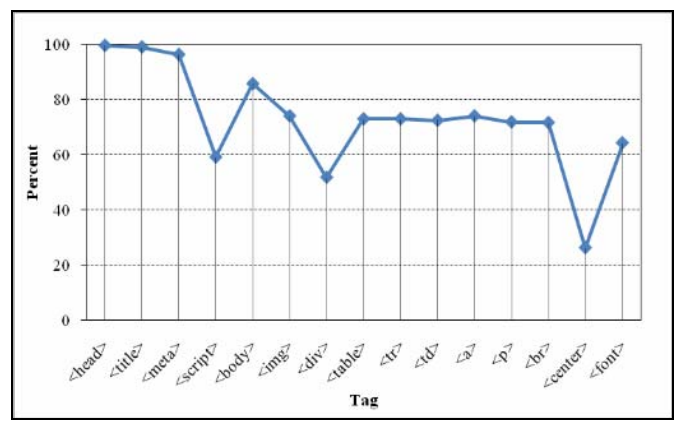

Figure 7. Percentage of tag usage for web documents

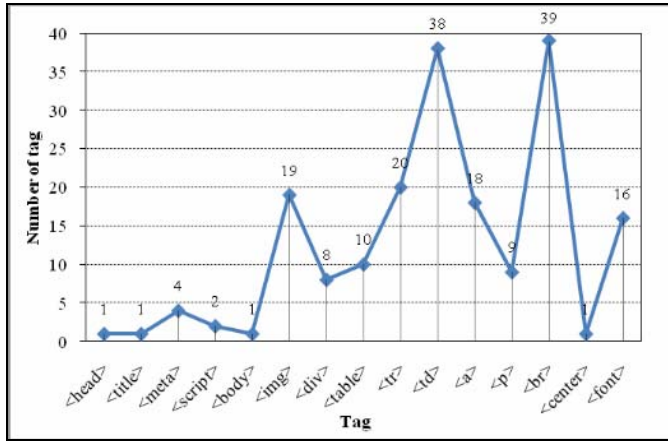

Figure 8. Amount of tag per document

In this research we have also studied the differences between file size for web documents. Generally, file size is related to the number of words that had been extracted. We discovered that the more the amount of words used, the bigger is the file size. This research showed that 57.52% of the file size consist of tag. Therefore, tagging usage is the main contributor to the file size. The file size for stopword consists of 12.24% while file size for text is 28.05%. The average file size of web documents is 0.0177 megabit and after going through the extracting process, the average file size shrinks to 0.0050 megabit. This research findings support Etzioni and Kosala who pointed out that websites emphasized on the presentation file on the web server compared to the content. Figure 9 shows the comparison of size file percentage for web documents.

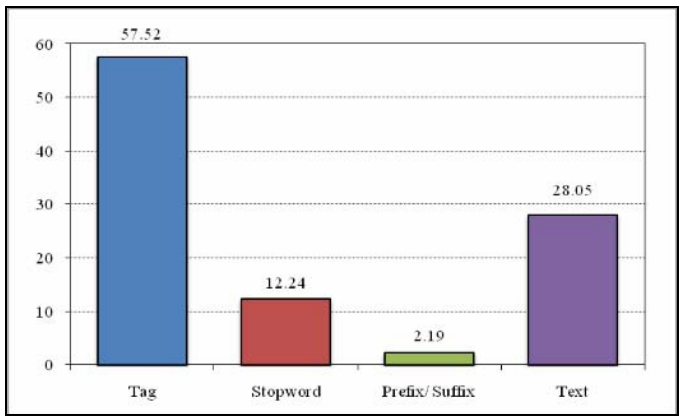

Figure 9. Percentage of file size for web documents

## VI. CONCLUSIONS

Our research findings showed that the website usage in companies on Bursa Malaysia is still minimal. This is because most of the companies have yet to realize the effects of their failure to attract Internet users' interests. Hence, local companies should be encouraged to move forward in their business using the Internet medium to avoid them from being left behind the competitors in the e-business market. However, it is also necessary for companies that already have websites to promote their websites to ensure that are able to reap the benefits of online business. Apart from that, the companies' website developers should be equipped with the awareness and sufficient knowledge in effective web presentation compared to the contents they want to highlight. This is to assure the companies websites are able to attract Internet users to visit them.

## REFERENCES

[1] Lawrence, S. and Giles, C. L., "Accessibility of Information on the Web" Nature, vol. 400, pp. 107-109, 1999.

[2] Pierre, J. M., On the automated classification of web sites, Electronic Transactions on Artificial Intelligence. http://www.ida.liu.se/ext/etai/ra/seweb/002/. [June 1, 2008], 2001.

[3] Gulli, A. and Signorini, A., "The indexable web is more than 11.5 billion pages", International World Wide Web Conference - Special interest tracks and posters of the 14th International Conference on World Wide Web, pp. 902-903, 2005.

[4] Internet Systems Consortium, ISC Internet Domain Survey, http://www.isc.org/index.pl, [January 19, 2008], 2008.

[5] Internet World Stat, Internet Usage World Stats - Internet and Population Statistics, http://www. internetworldstats.com/, [January 20, 2008], 2008.

[6] Lyman, P. and Hal R. V., How Much Information, http://www.sims.berkeley.edu/how-much-inf-2003, [June 1, 2009], 2003.

[7] Berners-Lee, T., Weaving the Web, London: Orion Business Books, 1999.

[8] Knuckles, C. D. and Yuen, D. S., Web Applications: Concepts and Real World Design, New Jersey: John Wiley & Sons, 2005.

[9] Baeza-Yates, R. and Ribeiro-Neto, B., Modern Information Retrieval, Essex: Addison Wesley, 1999.

[10] Etzioni, O. "The World Wide Web: quagmire or gold mine?", Communications of the ACM, vol. 39(11), pp. 65-68, 1996.

[11] Kosala, R. and Blockeel, H., "Web Mining Research: A Survey", ACM SIGKDD, vol 2(1), pp. 1-15, 2000.

[12] Ruhaya Atan, Nafsiah Mohamed and Normahiran Yatim, "E-Reporting of Corporate Financial Information. Web-based Financial Reporting in Malaysia", Seminar on Accounting and Information Technology, Convention Centre, Universiti Utara Malaysia, November 13-14, 2000.

[13] Noor Azizi Ismail and Mahamad Tayib, "Financial Reporting Disclosure on the Internet by Malaysian Public Listed Companies", Akauntan Nasional, vol. 13(10), pp. 28-33, 2000.

[14] Rosli Mohamad, Mudzamir Mohamed and Amdan Mohamed, "Internet Financial Reporting (IFR) in Malaysia: A Survey of Contents and Presentations", Accounting Seminar 2003, Putra Palace, Kangar, Perlis, Disember 8-10, 2003.

[15] Kamarulbaraini Keliwon and Khairul Azman Aziz, "Web financial reporting in Malaysia: The current stage", Proceedings of International Conference on E-Commerce 2005, pp. 59-65, 2005.

[16] The Star, SMIs still don't want to go online, pp. 52, 24/07/2001.

[17] Craven, T. C., "HTML tags as extraction cues for Web page description construction", Informing Science Journal, vol. 6, 1-12, 2003.

## AUTHORS PROFILE

**Dr Mohd Shahizan Othman** received his BSc in Computer Science with a major in Information Systems from Universiti Teknologi Malaysia (UTM), Malaysia, in 1998. Then he earned Msc in Information Technology from the Universiti Kebangsaan Malaysia (UKM), Malaysia. Soon after, he graduated for his PhD in Web Information Extraction, Information Retrieval and Machine Learning from UKM. He is currently a senior lecturer at the Faculty of Computer Science and Information Systems, UTM. His research interests are in information extraction and information retrieval on the web, web data mining, content management and machine learning.

**Lizawati Mi Yusuf** received her BSc in Computer Science from Universiti Teknologi Malaysia (UTM), Malaysia and Master degree of Information Technology with a major in Industrial Computing from Universiti Kebangsaan

Malaysia. She is currently pursuing PhD degree in Faculty of Computer Science and Information System, Universiti Teknologi Malaysia. Her fields of interest are high performance computing, numerical analysis, soft computing and information retrieval.

**Associate Professor Dr Juhana Salim** obtained a Bachelor in Library and Information Science (Universiti Technology of MARA), Bachelor of Art (Sociology) and Master of Science in Library Science from Western Michigan University and Ph.D (Information Science) from Universiti Kebangsaan M.alaysia. She worked as a Reference Librarian from June 1982 till December 1995. Her academic career started when she joined the Faculty of Information Science, Universiti Kebangsaan Malaysia in January 1996. She presented and published many papers in information skills, knowledge management and information organization. Her research on Web Resource Extraction and Gateway to Digital Library Libraries received gold and silver awards at local and International exhibitions. In 2004, she was awarded the Japan Society for the Promotion of Science travel grant to conduct research on cultural behavioural perspective on knowledge management.